\documentstyle[preprint,aps,fleqn]{revtex}
 
\tighten

\begin{document}
\draft
 
\title{Sum-Rule Inequalities and  a Toy Model Paradox} 
\author{ T.G. Steele, S. Alavian, J. Kwan }
\address{Department of Physics \& Engineering Physics and \\
Linear Accelerator Laboratory\\
University of Saskatchewan\\
Saskatoon, Saskatchewan S7N 0W0, Canada
}
\date{August 28, 1996}
\maketitle

\begin{abstract}
Fundamental inequalities for QCD sum-rules are applied to resolve a paradox
recently encountered in a sum-rule calculation \cite{piv}.  This paradox was encountered
in a toy model known to be free of resonances that yields an apparent
resonance using a standard sum-rule stability analysis. Application of the 
inequalities 
does not support the existence of a well defined sum-rule calculation, and
shows a strong distinction from typical behaviour in QCD.  
\end{abstract}

\vspace{0.2in}

QCD sum-rules\cite{svz,svz2,rry,nar} have demonstrated their utility in 
numerous theoretical determinations of hadronic properties.  In this
approach the QCD condensates  parametrize nonperturbative aspects
of the vacuum, and  
are an essential
feature of the sum-rules used to determine hadronic properties.

An apparent paradox in a sum-rule calculation has recently been encountered
\cite{piv}.  This paradox consists of a toy model known to be free of resonances
that yields an apparent resonance after the standard sum-rule stability method is applied
to the truncated correlation function.  
Clearly this raises concerns about the reliability of the stability criteria 
used to obtain predictions from QCD sum-rules.

Fundamental inequalities for sum-rules have recently been developed \cite{ineq}.
These inequalities must be satisfied if the sum-rule is  consistent with an 
integrated cross-section, but the analysis does not require detailed phenomenological
input on the nature of this cross-section.  These inequalities thus provide
 novel, valuable criteria for studying the validity and self-consistency 
of a sum-rule calculation.  QCD sum-rule inequalities 
are also of phenomenological value since they have been utilized to obtain bounds
related to the electromagnetic polarizability of charged pions \cite{polar}.

Laplace or Borel sum-rules are derived from a 
typical subtracted dispersion relation
\begin{equation}
-\frac{d\Pi(Q^2)}{dQ^2}=\frac{1}{\pi}\int\limits_{t_0}^{\infty}
\frac{ Im\Pi(t)}{\left(t+Q^2\right)^2}\, dt
\label{disp}
\end{equation}
where $t_0$ is  a physical threshold.  The Borel transform operator 
$\hat B$ \cite{svz} is then applied to the dispersion relation and the continuum
is subtracted leading to the Laplace sum-rules.
\begin{equation}
{\cal R}_k(\tau,s_0)=\frac{1}{\tau}\hat B\left[Q^2k\Pi(Q^2)\right]-{\rm continuum}
=\frac{1}{\pi}\int\limits_{t_0}^{s_0}t^ke^{-t\tau}Im\pi(t)\,dt
\label{sr}
\end{equation}
The parameter $s_0$ represents the continuum threshold leading to duality 
between the field theory and the physical process related to $Im\Pi(t)$.

H\"older's inequality\cite{hold,hold2} for integrals defined over a measure 
$d\mu$ is
\begin{eqnarray}
\biggl|\int_{t_1}^{t_2} f(t)g(t) d\mu \biggr| &\le& 
\left(\int_{t_1}^{t_2} \big|f(t)\big|^ p d\mu \right)^{1/p}
\left(\int_{t_1}^{t_2} \big|g(t)\big|^q d\mu \right)^{1/q}, \nonumber \\
&& \label{holder} \\[-1em] 
\frac{1}{p}+\frac{1}{q} &=&1~;\quad p,~q\ge 1\quad . \nonumber
\end{eqnarray}
When $p=q=2$ the H\"older inequality reduces to the well known Schwarz 
inequality. The key idea in applying H\"older's inequality to sum-rules is
recognizing that for a typical correlation function $\Pi(Q^2)$, 
$Im\, \Pi(t)$ is positive because of its relation to physical
cross-sections and can thus serve as the measure
$d\mu=Im\,\Pi(t) dt$ in (\ref{holder}).
Returning to (\ref{holder}) with $d\mu=Im\Pi(t) dt$, $f(t)=t^\alpha
e^{-at\tau}$, $g(t)=t^\beta e^{-bt\tau}$ and appropriate
integration limits we find
\begin{equation}
{\cal R}_{\alpha+\beta}(\tau,s_0)\le
{\cal R}^{1/p}_{\alpha p}(ap\tau, s_0)
{\cal R}^{1/q}_{\beta q}(bq\tau, s_0)
\;;\quad a+b=1\quad .
\label{rat1}
\end{equation}
Imposing restrictions that we have the integer values $k$ needed for
the sum-rules (\ref{sr}) leads to the following set of inequalities.
\begin{eqnarray}
{\cal R}_0[\omega\tau_{min}+(1-w)\tau_{max},s_0] &\le&
{\cal R}^\omega_0[\tau_{min},s_0] {\cal R}^{1-\omega}_0[\tau_{max},s_0],
\label{ineqa} \\
{\cal R}_1[\omega\tau_{min}+(1-w)\tau_{max},s_0] &\le&
{\cal R}^\omega_1[\tau_{min},s_0] {\cal R}^{1-\omega}_1[\tau_{max},s_0],
\label{ineqb}\\
{\cal R}_1[\frac{\tau_{min}+\tau_{max}}{2},s_0] &\le&
{\cal R}_2^{1/2}[\tau_{min},s_0]{\cal R}_0^{1/2}[\tau_{max},s_0],
\label{ineqc}\\
{\cal R}_1[\frac{\tau_{min}+\tau_{max}}{2},s_0] &\le&
{\cal R}_0^{1/2}[\tau_{min},s_0]{\cal R}_2^{1/2}[\tau_{max},s_0],
\label{ineqd}\\
0\le\omega\le 1~&;&\quad\tau_{min}\le\tau_{max}
\end{eqnarray}
Similar inequalities can be obtained for higher sum-rules with
$k\ge 2$.  Furthermore, for small $\delta\tau\equiv\tau_{max}-\tau_{min}$ 
(\ref{ineqc}) and (\ref{ineqd})
are in principle contained in the first two inequalities.

In summary, if the sum-rules are a valid and consistent representation of 
the integration of $Im\Pi(t)$ in (\ref{sr}) then  the sum-rules 
${\cal R}_k(\tau,s_0)$ must satisfy the following fundamental inequalities.
\begin{eqnarray}
\rho_0 &\equiv&\frac{{\cal R}_0[\tau+(1-
\omega)\delta\tau,s_0]}{{\cal R}_0^\omega[\tau,s_0]
{\cal R}_0^{1-\omega}[\tau+\delta\tau,s_0]} \le 1
\quad \forall ~0\le \omega \le 1
\label{rat2a}\\
\rho_1 &\equiv&\frac{{\cal R}_1[\omega\tau+(1-
\omega)\delta\tau,s_0]}{{\cal R}_1^\omega[\tau,s_0]
{\cal R}_1^{1-\omega}[\tau+\delta\tau,s_0]}\le 1 
\quad \forall ~0\le \omega \le 1
\label{rat2b}
\end{eqnarray}
Provided that $\delta\tau$ is reasonably small (in QCD $\delta\tau\approx 0.1\,GeV^{-2}$
appears sufficient)
these inequalities are  insensitive to the value of $\delta\tau$,
permitting a simple determination of the $\tau$, $s_0$ parameter space 
satisfying the inequalities.

In the standard stability analysis of QCD sum-rules, the ratio 
${\cal R}_1/{\cal R}_0$ gives the mass of the lightest resonance in the 
simple ``resonance plus continuum'' model for ${\rm Im}\,\Pi(t)$.  Thus if this stability
analysis is to be consistent with H\"older's inequalities then (\ref{rat2a}) and
(\ref{rat2b}) should be satisfied in the {\em same} region of $\tau$, $s_0$ parameter
space in which the stability analysis occurs.
 The $s_0$, $\tau$ parameter space consistent with the inequalities (\ref{rat2a},\ref{rat2b}) 
 is shown in
 Figures 1-3 for a wide variety of QCD sum-rule applications
ranging from light to heavy quark systems \cite{svz,svz2,rry,nar,bagan}
 Also shown in the figures is the
 the range of $\tau$, $s_0$ resulting from the corresponding stability analysis
 \cite{svz,svz2,rry,nar,bagan}.  
In all cases, a large portion of the stability region is seen 
to be consistent with the inequalities, indicating a reliable sum-rule analysis.
The effect of uncertainties intrinsic to the sum-rule method has been analyzed in
\cite{ineq}, and the result is a slight increase in the area of the parameter space
consistent with the inequalities.

The allowed sum-rule
parameter space in Figures 1-3 exhibits two crucial features that we believe are characteristic of QCD:
a failure of duality below a critical value of $s_0$, and an upper limit on $\tau$ corresponding
to a lower bound on the energy.
It is also interesting to note that the narrow-width single resonance approximation
provides a phenomenological contribution on the right-hand side of (\ref{sr}) 
given by $F^2e^{-M^2\tau}$.  This satisfies the {\em equality} in (\ref{rat2a},\ref{rat2b}),
and so the borders of the parameter space in Figures 1-3 represents points where the 
sum-rule is in good agreement with a single (narrow-width) resonance model.  This provides
a practical criteria for choosing the optimal $s_0$ in a sum-rule analysis: a value of $s_0$ which corresponds to a wide, flat
border of the allowed parameter space has maximum agreement with a 
single narrow-width resonance.  As is seen from consideration of the values of $s_0$
used in the actual sum-rule analyses (horizontal lines in Figures 1-3) 
\cite{svz,svz2,rry,nar,bagan}
this criteria is consistent with the range of $s_0$ corresponding to the 
conventional stability analysis.

We now consider the application of the inequalities to the toy model of
\cite{piv} where a correlator of composite operators is calculated in the ladder 
approximation for a $\phi^3$ theory.  The 
spectral density in this model has a non-physical behaviour in some momentum
regions where the ladder diagrams do not accurately represent the exact correlation function.
The following sum-rules can be obtained from
a truncated version of the correlator:
\footnote{The notation  $\tau=1/M^2$ is used in \cite{piv}.}
\begin{eqnarray}
{\cal R}_0(\tau,s_0)=\frac{1-e^{-s_0\tau}}{\tau}-6\zeta(3) +20\zeta(5)\tau
-35\zeta(7)\tau^2+42\zeta(9)\tau^3+\ldots
\label{toy1}\\
{\cal R}_1(\tau,s_0)=\frac{ 1-\left(1+s_0\tau\right)e^{-s_0\tau}}{\tau^2}
-20\zeta(5)+70\zeta(7)\tau-129\zeta(9)\tau^2+\ldots
\label{toy2}
\end{eqnarray}
Despite the non-physical aspects of this model, when the truncated correlator is analysed with 
conventional sum-rule stability techniques, an apparently physical resonance and 
continuum of positive spectral density is discovered in contradiction with the non-truncated
ladder approximation \cite{piv}.  This clearly raises concerns about the reliability  
of conventional QCD sum-rule stability techniques where information on
physical resonances is extracted from  truncated  versions of QCD correlation functions.

The $s_0$, $\tau$ parameter space where the toy model (\ref{toy1},\ref{toy2}) is consistent with the 
inequalities (\ref{rat2a},\ref{rat2b})  is shown in Figure 4.  
Also shown is a boxed region corresponding to the range of $\tau$ and $s_0$
obtained from the stability analysis \cite{piv} (an uncertainty for $s_0$
was not given in \cite{piv}, so a 10\% deviation has been assumed 
for the quoted optimal value of $s_0=16.6$ at the lower bound of the boxed region).  
In contrast to the QCD examples, there
is no overlap between the stability and inequality parameter space, indicating an 
inconsistency with the fundamental inequalities.  The parameter  space consistent with the inequalities 
bears little resemblance to the QCD examples since the flat border 
of the parameter space needed for maximal agreement with a single narrow resonance is absent.

Thus  in contrast with the conventional sum-rule stability techniques which identify a spurious
resonance and physical continuum in the toy model, 
application of  fundamental sum-rule  inequalities reveals that this toy model
does not provide a convincing description of a resonance and physical continuum.  Furthermore,    
the inequalities provide a clear distinction between the toy model and QCD.
This indicates the importance
of supplementing the traditional sum-rule stability analysis with 
the fundamental inequalities developed in \cite{ineq}.
 
\bigskip\noindent
{\bf Acknowledgements:}  
TGS and JK are grateful for the financial support of the Natural Sciences 
and Engineering Research Council of Canada (NSERC).  
We are grateful to M. Lavelle for  bringing reference 
\cite{piv} to our attention.  
TGS dedicates this work to the memory of G. Fitzhenry.

\newpage

\begin{figure}
\caption{
The shaded area represents the  region 
in the $s_0$, $\tau$ parameter space consistent with the
inequalities for the light-quark vector 
sum-rule related to the $\rho$ meson \protect\cite{svz,svz2,rry,nar}. 
The boxed region indicates the values of $\tau$, $s_0$ used in the sum-rule
stability analysis.  As determined in \protect\cite{lein}, a 25\% uncertainty 
in $s_0$ has been used.
}
\label{figure1}
\end{figure}

\begin{figure}
\caption{ 
Same as in Figure \protect\ref{figure1} except for the light quark axial-vector 
sum-rule related to the $A_1$ meson \protect\cite{svz,svz2,rry,nar}. 
}
\label{figure2}
\end{figure} 

\begin{figure}
\caption{
Same as Figure \protect\ref{figure1} except for the heavy quark sum-rule
related to the $B$ meson \protect\cite{bagan}.  The convention for $s_0$ and $\tau$ 
differs
from Figures \protect\ref{figure1} and \protect\ref{figure2} for consistency with \protect\cite{bagan} which takes into account the relation with
heavy quark effective theories.  In particular $\tau=1/\Delta M$ in the notation of
\protect\cite{bagan}.
}
\label{figure3}
\end{figure}

\begin{figure}
\caption{
Shaded region indicates the
parameter space consistent with the inequalities for  the sum-rule in the toy model
\protect\cite{piv}.
The value $\delta\tau\approx 0.001$ has chosen because of the
lower scales for $\tau$ relevant to the toy model.
The boxed region indicates the values of $\tau$, $s_0$ used in the sum-rule
stability analysis \protect\cite{piv}.
}
\label{figure4}
\end{figure}

\end{document}